\documentstyle[12pt,epsfig,color]{article}
 \newcommand{\be}{\begin{eqnarray}}
 \newcommand{\ee}{\end{eqnarray}}
 \newcommand{\beq}{\begin{equation}}
 \newcommand{\eeq}{\end{equation}}
 \newcommand{\ba}{\begin{array}{1}}
 \newcommand{\ea}{\end{array}}
 \newcommand{\bb}{\begin{thebibliography}}
 \newcommand{\eb}{\end{thebibliography}}

\begin{document}
\begin{center}
{\bfseries Dielectron production in pion-nucleon reactions and
  form factor of baryon transition within the time-like region
}
\vskip 5mm
A.P.Jerusalimov, G.I.Lykasov
\vskip 5mm
{\small {\it Joint Institute for
Nuclear Research, 141980 Dubna, Russia}} \\
\end{center}

\vskip 5mm
\centerline{\bf Abstract}

Dielectron production in reactions $\pi^- p \rightarrow n e^+e^-$
and $\pi^- p \rightarrow n e^+e^- \gamma$  at energies less than 1 GeV is studied assuming
electron-positron pair production to occur in the virtual time-like photon splitting process.
Theoretical predictions of the effective mass distribution of dielectrons
and their angular dependence are presented. Extraction of the electromagnetic form factor of
baryon transition in the time-like region from future experiments of the HADES Collaboration
is discussed. 

\vskip 10mm
\section{\label{sec:intro}Introduction}

Investigation of the electromagnetic form factor ($FF$) of hadrons provides significant
information about their structure.
For example, measurement of the $e^+e^-\rightarrow\pi^+\pi^-$ cross section results in the pion $FF$ in the time 
like region, which results in parameters of the $\rho$-meson and its excited states. This cross section has
been measured in Orsay \cite{Orsay:1968}, Novosibirsk \cite{Novosibirsk:1985}-\cite{Novosib:2007}, Frascati (KLEO) 
\cite{KLOE:2013}, SLAC (BABAR) \cite{BABAR:2009,BABAR:2013} and Beijing (BESIII) \cite{BESIII:2016}. 

There is another way to investigate the electromagnetic form factor in time-like four-momentum space 
in processes of $e^+e^-$-pair production in hadron-hadron, hadron-nucleus and nucleus-nucleus 
collisions with the High Acceptance Di-Electron Spectrometer (HADES) \cite{HADES:2009}. Experimental
\cite{HADES:2010,HADES:2017} and theoretical \cite{Gaffen:1962}-\cite{JL:2017}
analysis of dielectron production in $pp,np,pA$ and $AA$ collisions leads to detailed 
information on the reaction mechanism, which is due mainly to the creation of baryonic and mesonic resonances 
in the intermediate state decayed into $e^+e^-$-pair. 
The study of exclusive dielectron production
in meson-nucleon interaction at not large initial energies, for example, at about a few hundred MeV simplifies
the theoretical analysis because the number of baryoinic and mesonic resonances in the intermediate 
state is not large. In this case the $e^+e^-$ production in $\pi N$ interaction can be studied assuming the electron-positron
pair to be produced in the virtual time-like photon splitting
and to be considered a $e^+e^-$ dipole. From future experimental data on
$\pi^- p\rightarrow\gamma^* n\rightarrow e^+e^- n$ planned
at HADES \cite{HADES:2019} interesting information about this $e^+e^-$  dipole, for example, its form factor
can be found. In this paper we continue our previous theoretical analysis of the reaction
$\pi^- p\rightarrow e^+e^- n$ \cite{JL:2018}
at intermediate energies less than 1 $GeV$. In addition to that we 
include the contributions of a few additional channels, namely,
$\pi^- p \rightarrow\pi^0 n\rightarrow e^+e^- \gamma n$,
$\pi^- p \rightarrow\rho^0 n\rightarrow e^+e^- n$ and $\pi^- p \rightarrow\eta^0 n\rightarrow e^+e^- \gamma n$,
which increase the dielectron effective mass distribution and result in predictions of the
future HADES experiment for the form factor of the $e^+e^-$ dipole.      
            `
\section{General formalism}

\begin{figure}[ht]
{\epsfig{file= 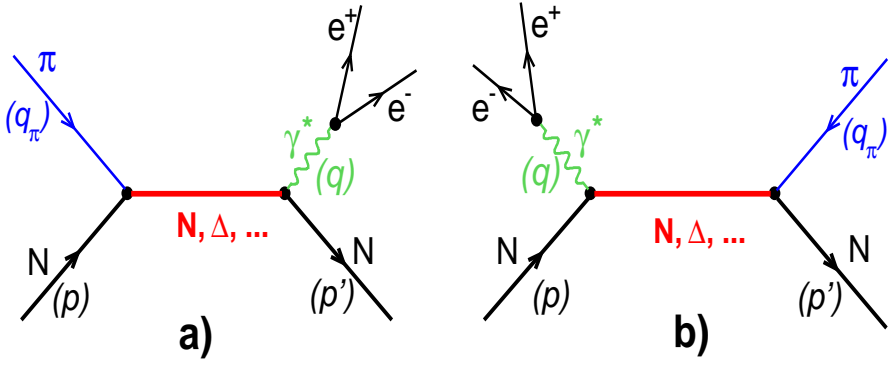,height=5.cm,width=12.cm  }}
\caption[Fig.1]{The one-nucleon or one-nucleon resonance exchange graph in the $s$-channel 
(a) and $u$-channel (b) of the IPE $\pi N\rightarrow\gamma N\rightarrow e^+e^- N$ process.}
\label{Fig.1}
\end{figure}

\begin{figure}[ht]
{\epsfig{file= 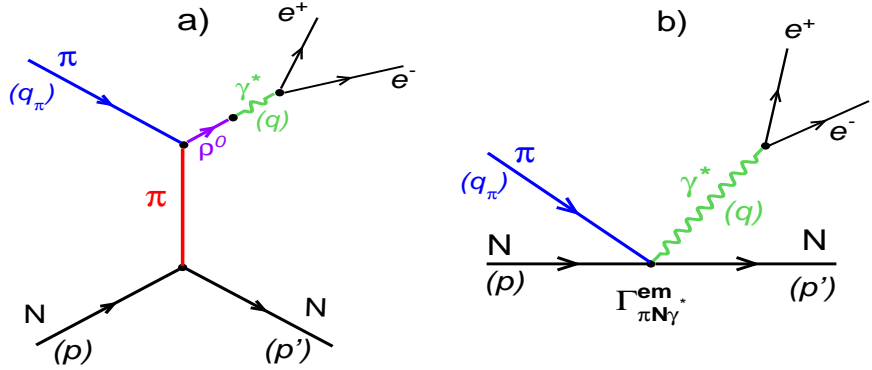,height=5.cm,width=12.cm  }}
\caption[Fig.2]{The one-pion exchange graph in the $t$-channel (a)
and the electromagnetic contact term (b) for the IPE.}
\label{Fig.2}
\end{figure}

\begin{figure}[ht]
{\epsfig{file= 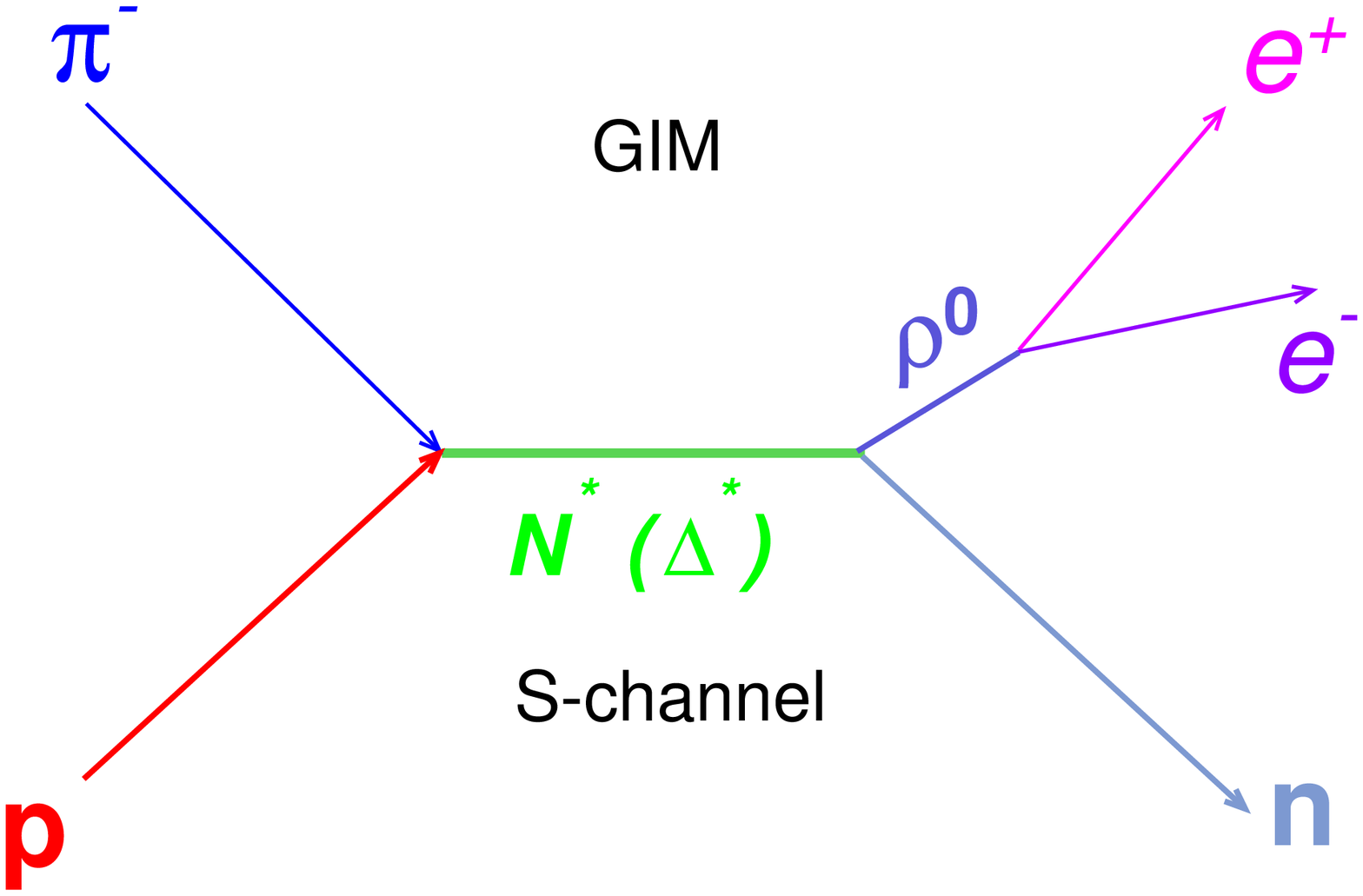,height=5.cm,width=12.cm  }}
{\epsfig{file= 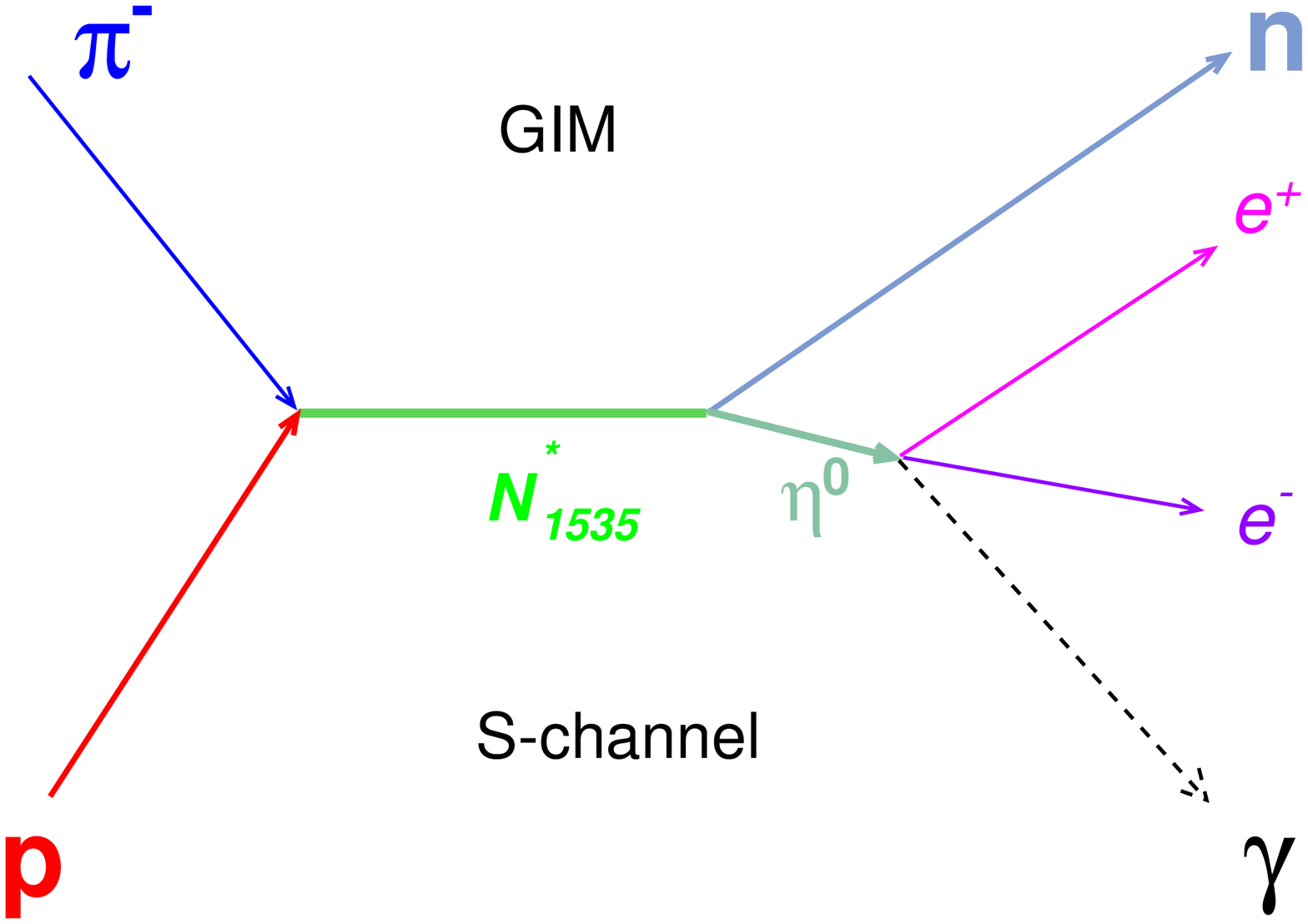,height=5.cm,width=12.cm  }}
\caption[Fig.1]{The one-nucleon resonance exchange graph in the $s$-channel of
the IPE with production of $\rho^0\rightarrow e^+e^-$ (top) and $\eta^0\rightarrow e^+e^- \gamma $ (bottom). } 
\label{Fig.3}
\end{figure}

We analyze the reaction $\pi N\rightarrow\gamma^* N\rightarrow e^+e^- N$ within the
unified model. \textcolor{black}{This} means that in the one-photon approximation,
owing to $T$-invariance, three reactions $\gamma N\rightarrow\pi N,
e N\rightarrow e\pi N$ and $\pi N\rightarrow e^+e^- N$ are related to the process 
$\gamma^* N\leftrightarrow\pi N$ \textcolor{black}{by} the hadron current
$J_\mu(s,t,m_\gamma^2)$, where $m_\gamma^2=0,>0$ and $m_\gamma^2<0$ correspond 
\textcolor{black}{to} pion photoproduction, electroproduction and inverse pion
electroproduction (IPE), respectively \cite{Berends:1967,Surovtsev:2005}.
In our previous paper \cite{JL:2018} application of this unified model to calculate
the $e^+e^-$ effective mass distribution and the angular distribution of the virtual photon 
decayed into $e^+e^-$ in the IPE processes was presented in detail. A satisfactory 
description of the data at a pion initial momentum of
about 300 MeV/c was presented. Therefore, we shall omit the details
of the matrix element calculation for reaction
$\pi^- p\rightarrow\gamma^* n\rightarrow e^+e^- n$ and 
only present the general forms for the  
effective mass distribution $d\sigma/dM_{e^+e^-}$ of the $e^+ e^-$ pair and the
angular distribution $d\sigma/d\cos\theta_{\gamma^*}^*$ 
of the virtual $\gamma^*$ in the $\pi-N$ c.m.s.\cite{JL:2018}:
\begin{equation}
\frac{d\sigma}{d M_{e^+e^-}}~=~\frac{\alpha_{em}^2}{12\pi M_{e^+e^-}s\lambda^{1/2}(s,\mu_\pi^2,m^2)}
\int_{t^-}^{t^+}\frac{\lambda^{1/2}(s,q^2,m^2)}{s}(FF(q^2))^2\sum_{spins}J_\mu J_\nu^+ dt~;
\label{def:effmassdist}
 \end{equation} 
 \begin{equation}
\frac{d\sigma}{d \cos\theta_{\gamma^*}^*}~=~\frac{\alpha_{em}^2}{48\pi s\lambda^{1/2}(s,\mu_\pi^2,m^2)}
\int\frac{\lambda^{1/2}(s,q^2,m^2)}{q^2}(FF(q^2))^2\sum_{spins}J_\mu J_\nu^+dq^2~,
\label{def:costetadist}
 \end{equation} 
where $\alpha_{em}=e^2/4\pi=1/137$; $ \lambda(x^2,y^2,z^2)=(x^2-(y+z)^2) (x^2-(y-z)^2)$ and 
$\sum_{spins}J_\mu J_\nu^+=W_{\mu\nu}$ is the hadronic tensor, which was calculated in \cite{JL:2018}
including the graphs of Figs.~(\ref{Fig.1},\ref{Fig.2}).
Let us note that the contribution of the $t$-channel one exchange graph Fig.~(\ref{Fig.2}a) to $d\sigma/dM_{ee}$ 
is very small at initial momenta less than 1GeV$/$c, see, for example, \cite{pipiN} and
references therein,
compared to contributions of the one baryonic exchange in the $s$-channel (Fig.~(\ref{Fig.1}a))
and in $u$-channel (Fig.(\ref{Fig.1}b)).        
At initial pion momenta of about 300 MeV$/$c the graphs of Fig.~(\ref{Fig.1}) with the
$\Delta$-isobar
exchange result in the main contribution to $d\sigma/d M_{e^+e^-}$. At higher initial momenta up
to 700-800 MeV$/$c, which correspond to the HADES experiment with the pion beam \cite{HADES:2019}, the
additional graphs corresponding to the production of $\rho^0$ and $\eta^0$, can contribute to the
matrix element of this reaction. They are presented in Fig.~(\ref{Fig.3}).  
The contributions of these graphs with the exchange of different baryonic
resonances were calculated
within the Generalized Isobar Model (GIM)~\cite{GIM:1975,GIM:1992}, see APPENDIX. 

\section{Electromagnetic form factor at the time-like region}

As it was mentioned above, the pion form factor $F_\pi$ in the time-like
region was measured directly in the annihilation process 
$e^+e^-\rightarrow\pi^+\pi^-$ from its cross section. Then, the mean value of the pion radius square was determined
from $F_\pi$ as follows \cite{Novosibirsk:1985,BESIII:2016}:
\begin{equation}
<r^2_\pi>~=~6\frac{dF_\pi(s)}{ds}|_{s=0},
\label{def:rpisq} 
\end{equation} 
where $s$ is the square of the initial energy in the c.m.s of $e^+e^-$. According to \cite{Novosibirsk:1985}, 
$<r^2_\pi>=$0.422$\pm$0.003$\pm$0.0013 fm$^2$.
The mean square radius of the charged $\rho$-meson, which can be determined, for example,
from its decay constant \cite{Krutov:2016}, is, about 0.56-0.6 fm$^2$, i.e., larger than the similar value
for a pion.
   
To include the virtual photon off-shellness in our process
$\pi^- p\rightarrow\gamma^* n\rightarrow e^+e^- n$, 
which can be large, up to about a few hundred MeV, we introduced the electromagnetic 
form factor $FF(q^2)$ in Eqs.~(\ref{def:effmassdist},\ref{def:costetadist}). We choose this FF in
two forms:
 \begin{equation}
FF(q^2)~=~\frac{\Lambda^2}{\Lambda^2 - q^2}
\label{def:monopFF} 
 \end{equation} 
and
 \begin{equation}
FF(q^2)~=~\exp(R^2q^2)
\label{def:exponFF} 
 \end{equation} 
 For the virtual photon in the time-like region, when $q^2>0$, the form factor $FF$ can
 become larger than 1, 
 therefore, it increases the $M_{ee}$-spectrum at large $q^2=M_{ee}^2$. Actually, the $FF$ in the form of
 Eq.(\ref{def:monopFF}) is similar to the form factor 
corresponding to the vector meson dominance model (VMD), when the parameter $\Lambda$ is the vector meson mass $M_V$.
In order to avoid the divergence in Eq.(\ref{def:monopFF}) at $q^2=\Lambda^2$ 
we also use the exponential form of $FF$ (Eq.~(\ref{def:exponFF})). 
The $e^+e^-$ pair in the $s$-channel (Fig.~(\ref{Fig.1})) is produced  from the baryonic transition in the
time-like region, and the form factor $FF(q^2)$ can determine the size of this region.
This form factor $FF$ can be 
extracted from future HADES experimental data on $d\sigma/d M_{e^+e^-}$ and
$d\sigma/d\cos\theta_{\gamma^*}^*$ in a similar way, as it was done for the pion $FF$ from
the $e^+e^-\rightarrow\pi^+\pi^-$ cross section 
\cite{Novosibirsk:1985,BESIII:2016}.
The mean value of the square radius of the electromagnetic baryonic transition (BT) is related to
the derivative of $FF$ like in Eq.~(\ref{def:rpisq})
\begin{equation}
<r^2_{BT}>~=~6\frac{dFF}{dq^2}|_{q^2=0}
\label{def:rdsq} 
\end{equation} 
So, knowing the $FF$ from future experiments on $\pi^- p\rightarrow e^+e^- n$
one can estimate the size of the time-like baryonic transition region.  

\section{Results and discussion}

\begin{figure}[ht]
{\epsfig{file=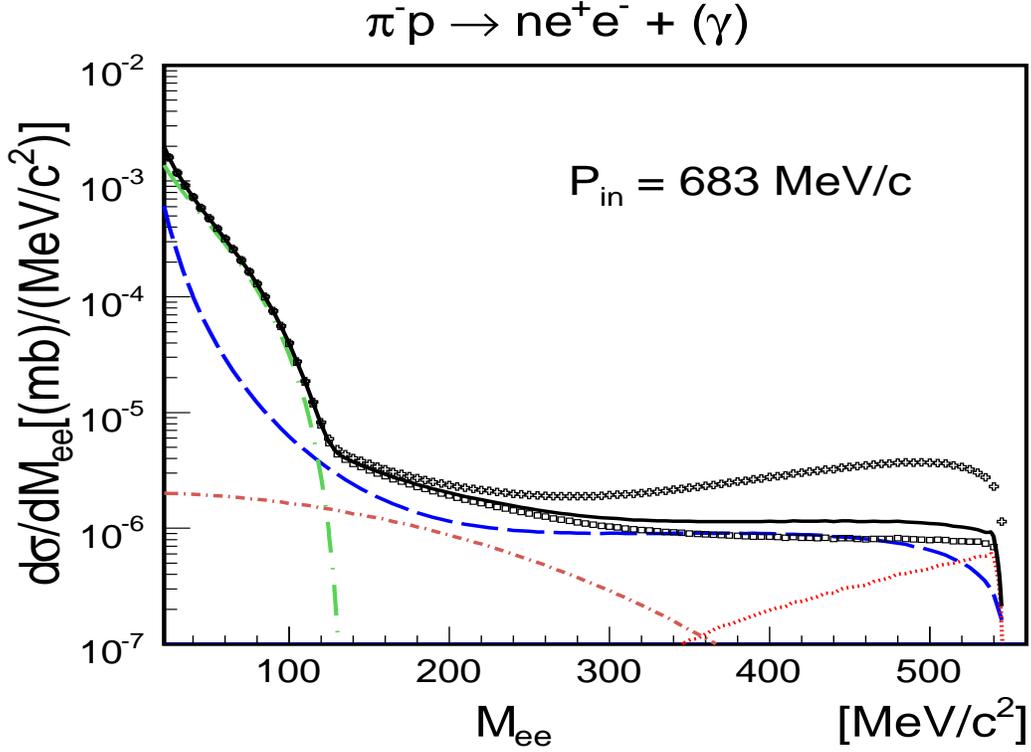,height=10.cm,width=14.cm  }}
\caption{Invariant mass distribution for 
dielectrons, $d\sigma/dM_{e^+e^-}$\textcolor{red}{,}  
produced in the reaction $\pi^- p\rightarrow e^+e^- n (\gamma)$. 
The solid curve corresponds to our total calculation including the exponential form
factor in the form of Eq.~\ref{def:exponFF} at $R=$ 1.6 (GeV/c)$^{-1}$); the
open circles correspond to the same calculation but at $R=$ 0 and the crosses
are our total calculation at $R=$ 3 GeV/c)$^{-1}$. The separate contributions to
this spectrum are presented by the blue long dashed line (the channel 
$\pi^- p\rightarrow e^+e^- n$), the green long dashed-dotted line 
($\pi^- p\rightarrow n\pi^0\rightarrow n e^+e^-\gamma$). the red
short dashed-dotted curve (the channel $\pi^- p\rightarrow n\eta^0\rightarrow n e^+e^-\gamma$,
Fig.~(\ref{Fig.3} bottom)) and the red dotted line (the channel 
$\pi^- p\rightarrow n\rho^0\rightarrow n e^+e^-$, Fig.~(\ref{Fig.3} top).
}
\label{Fig.4}
\end{figure}

\begin{figure}[ht]
{\epsfig{file=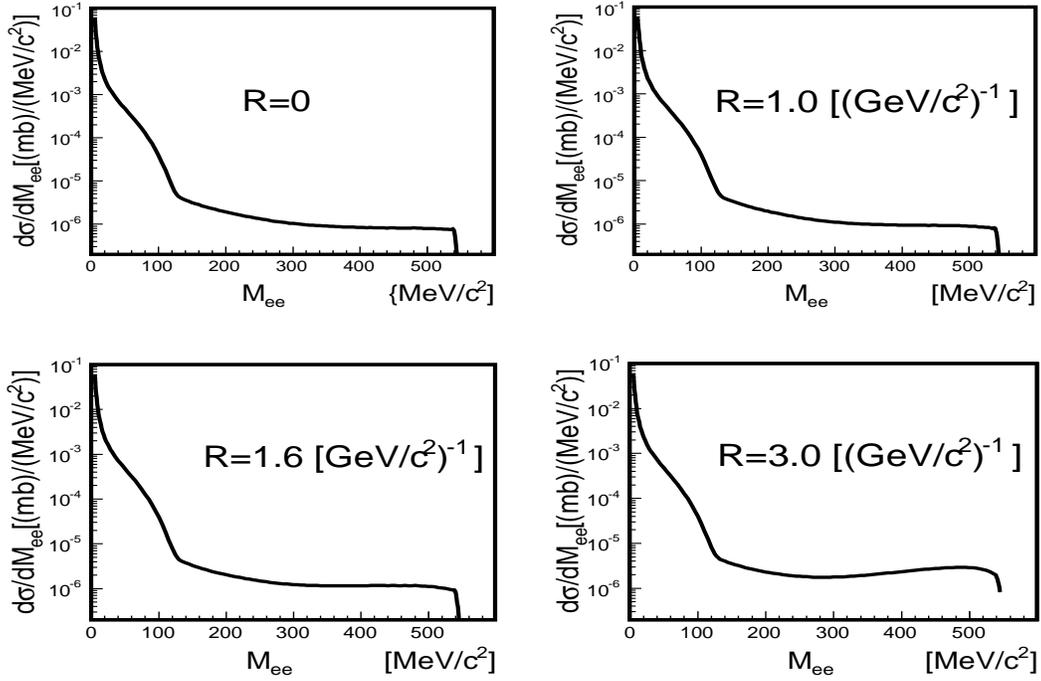,height=10.cm,width=14.cm  }}
\caption{Invariant mass distribution for 
dielectrons, $d\sigma/dM_{e^+e^-}$\textcolor{red}{,}  
produced in the reaction $\pi^- p\rightarrow e^+e^- n$ calculated by
including the form factor $FF$ in the form of Eq.~\ref{def:exponFF} 
at different values of parameter $R$. 
}
\label{Fig.5}
\end{figure}

We have calculated the distributions of $e^+e^-$ pairs as
functions of the dielectron effective mass $M_{e^+e^-}$ and
$cos\theta_{\gamma^*}$ for the processes $\pi^- p\rightarrow\gamma^* n \rightarrow e^+e^- n$
and $\pi^- p\rightarrow\gamma^* n \rightarrow e^+e^- n\gamma $ at initial pion momenta from a 
few hundred MeV$/$c up to 1 GeV$/$c. In Fig.~(\ref{Fig.4}) we present the distribution 
$d\sigma/dM_{ee}$ for these processes denoted as $\pi^- p\rightarrow\gamma^* n \rightarrow e^+e^- n (\gamma)$
at the initial pion momentum $P_{in}=$683 MeV/c corresponding to the HADES experiment.
The blue long dashed line in Fig.~(\ref{Fig.4}) corresponds to 
reaction $\pi^- p\rightarrow\gamma^* n \rightarrow e^+e^- n$ inputting the form factor $FF=$1; 
the green long dashed-dotted curve corresponds to the channel $\pi^- p\rightarrow\pi^0 n \rightarrow e^+e^-\gamma n$;
the red short dashed-dotted line and the red dotted curves correspond to the contributions of channels 
$\pi^- p\rightarrow n\eta^0\rightarrow e^+e^-\gamma n$ and $\pi^- p\rightarrow \rho^0 n\rightarrow e^+e^- n$, respectively. 
At the HADES facility it is very difficult to distinguish the channel $\pi^- p\rightarrow e^+e^- n$ from
the channel $\pi^- p\rightarrow e^+e^-\gamma n$, therefore, we incoherently sum up the
contributions of these four channels presented by open black circles in Fig.~(\ref{Fig.4}) inputting $FF=$ 1.
The $M_{ee}$-spectrum for this sum is denoted as the spectrum of the process $\pi^- p\rightarrow e^+e^- n (\gamma)$. 
Then, we investigate the sensitivity of our results to the form factor $FF$ chosen in the Gaussian form given by 
Eq.~(\ref{def:exponFF}). The solid line in Fig.~(\ref{Fig.4}) is the total $M_{ee}$-spectrum including all these channels
and inputting in the form factor $FF$ the parameter $R=$1.6 (GeV$/$c)$^{-1}=$ 0.32 fm., which corresponds to the
square $\rho$-meson radius of about 0.614 fm$^2$ \cite{Krutov:2016}. The crosses in
Fig.~(\ref{Fig.4}) are our calculations of the spectrum for the parameter
$R=$ 3 (GeV$/$c)$^{-1}$, which corresponds to the square radius $<r^2_{BT}>$ of
about 2 fm$^2$. Let us note that the proton charge radius $R_E\simeq 0.841$ fm. and
the proton magnetic radius $R_M=0.87$ fm. and the Zemach radius $R_Z=1.082$ fm., which reflects the spatial
distribution of magnetic moments smeared out by the charge distribution of the proton \cite{Zemch_FF}.
The sensitivity of the total  $M_{ee}$-spectrum to the parameter $R$ was presented in detail
in Fig.~(\ref{Fig.5}). One can see an enhancement in the spectrum at $M_{ee}>$ 300 MeV$/$c$^2$ at large values of the
parameter $R$. This enhancement could 
be due to a big off-shellness of the virtual photon $\gamma$. An excess of this spectrum over our
calculations performed without the form factor $FF$ can provide information on the form factor $FF$ of
electromagnetic baryon transition. Therefore, the future exclusive HADES experiments on the dielectron production in
pion-proton and pion-nucleus interactions one can permit to estimate the the size of the
time-like baryon transition to $e^+e^- N$, according to Eq.~(\ref{def:rdsq}).

\begin{figure}[ht]
{\epsfig{file=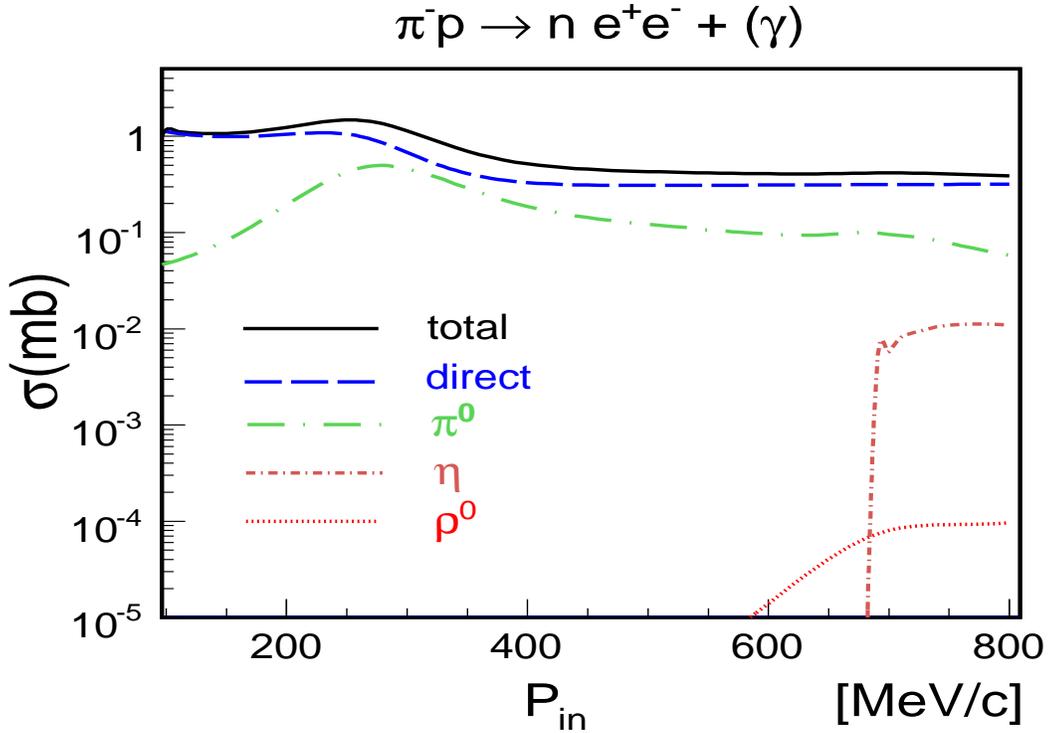,height=10.cm,width=14.cm  }}
\caption{The cross section of the process $\pi^- p\rightarrow e^+e^- n (\gamma)$ as a function of the
  initial pion momentum.
The notations are the same as in Fig.~\ref{Fig.4} and the calculations were performed at $R=$ 1.6 (GeV$/$c)$^{-1}=$ 0.32 fm.  
}
\label{Fig.6}
\end{figure}

In Fig.~(\ref{Fig.6}) we present the cross section of the process $\pi^- p\rightarrow e^+e^- n (\gamma)$ as a function
of the initial pion momentum. The notations are the same as in Fig.~(\ref{Fig.4}) and the calculations
were performed at $R=$ 1.6 (GeV$/$c)$^{-1}=$ 0.32 fm.
       
\begin{figure}[ht]
{\epsfig{file=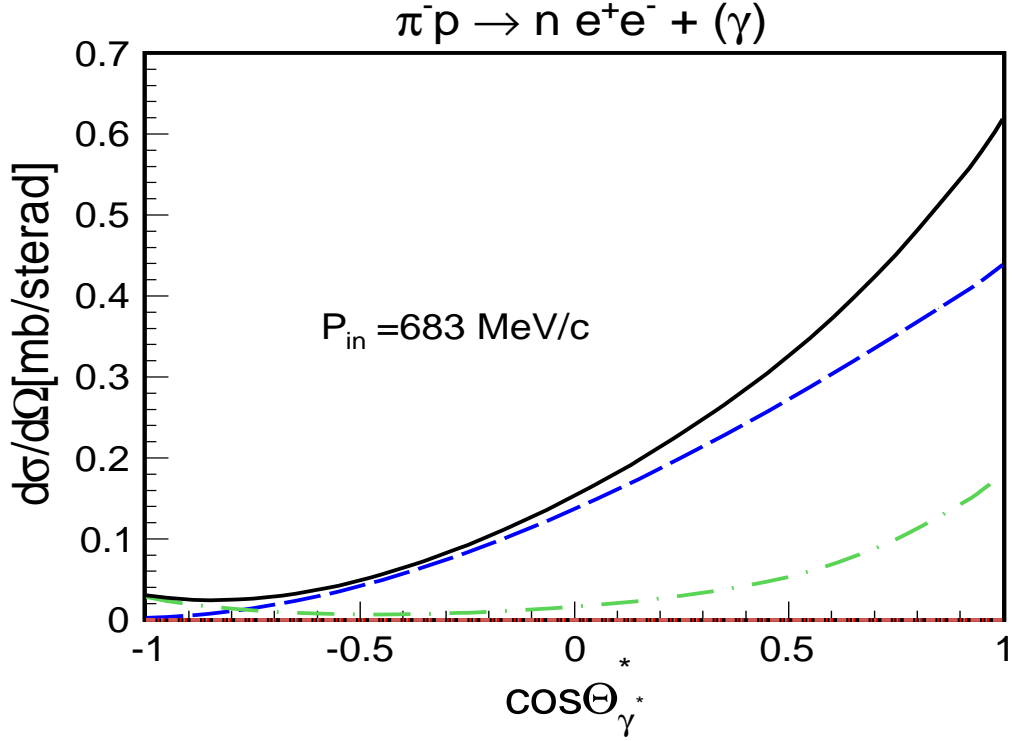,height=10.cm,width=14.cm}}
\caption{Angular distribution $d\sigma/d\cos\theta_{\gamma^*}$, 
where $\theta_{\gamma^*}$ is the angle of the virtual photon $\gamma^*$.  
The calculations were performed at $P_{in}=$  683 MeV$/$c and $R=$ 1.6 (GeV$^{-1}=$) 0.32 fm.   
} 
\label{Fig.7}
\end{figure}

In Fig.~(\ref{Fig.7}) the angular distribution $d\sigma/d\cos\theta_{\gamma^*}$
is presented at $P_{in}=$683 MeV/c and  $R=$ 1.6 (GeV$^{-1}=$) 0.32 fm. integrated over $q^2=M^2_{ee}$
As our calculations show, this distribution is practically not sensitive to the 
inclusion of $FF$. 
One can see the asymmetry of distribution $d\sigma/d\cos\theta_{\gamma^*}$, which depends on the initial momentum
very weakly, according to our calculations.   

\begin{figure}[ht]
{\epsfig{file=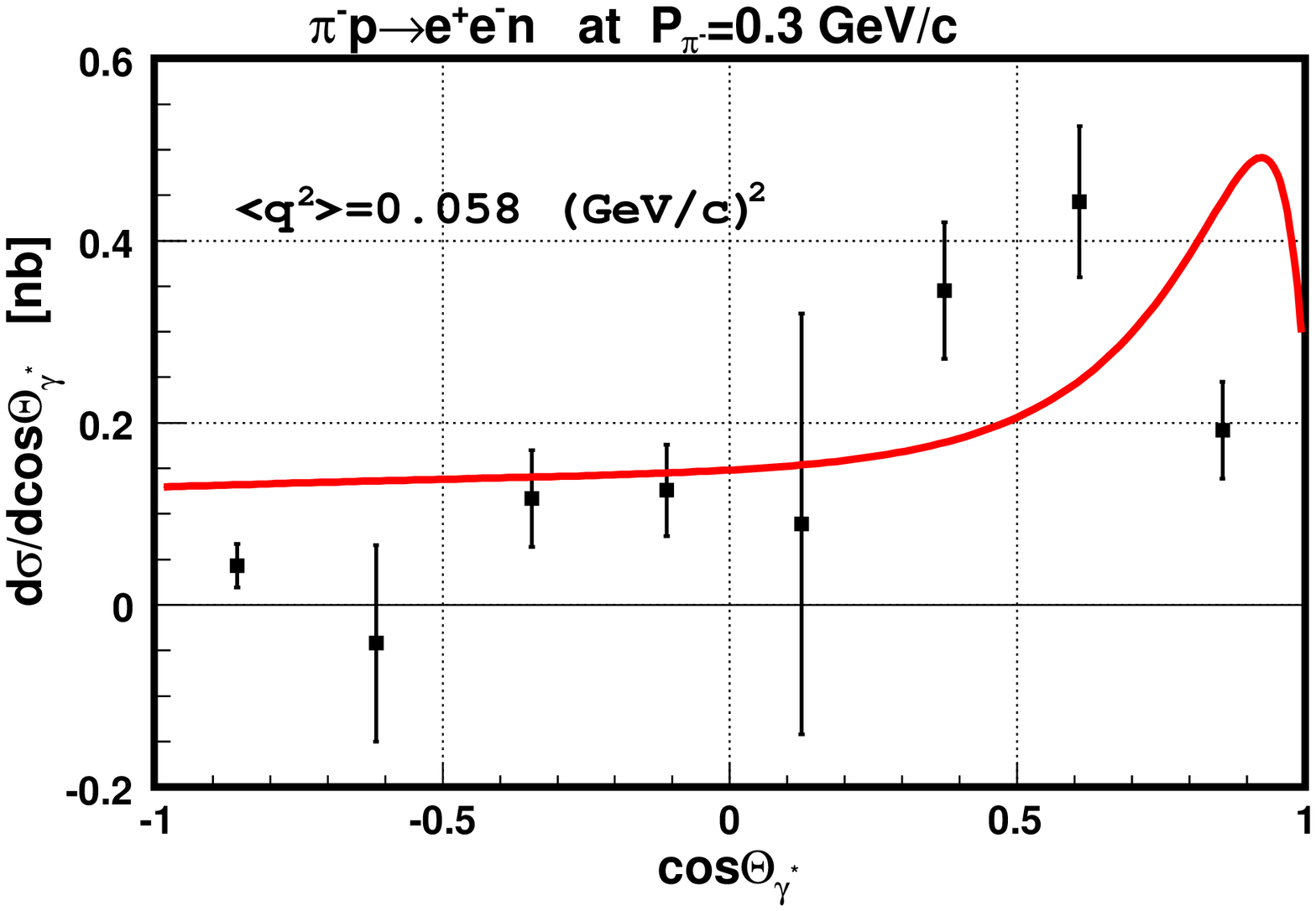,height=7.cm,width=7.2cm  }}
{\epsfig{file=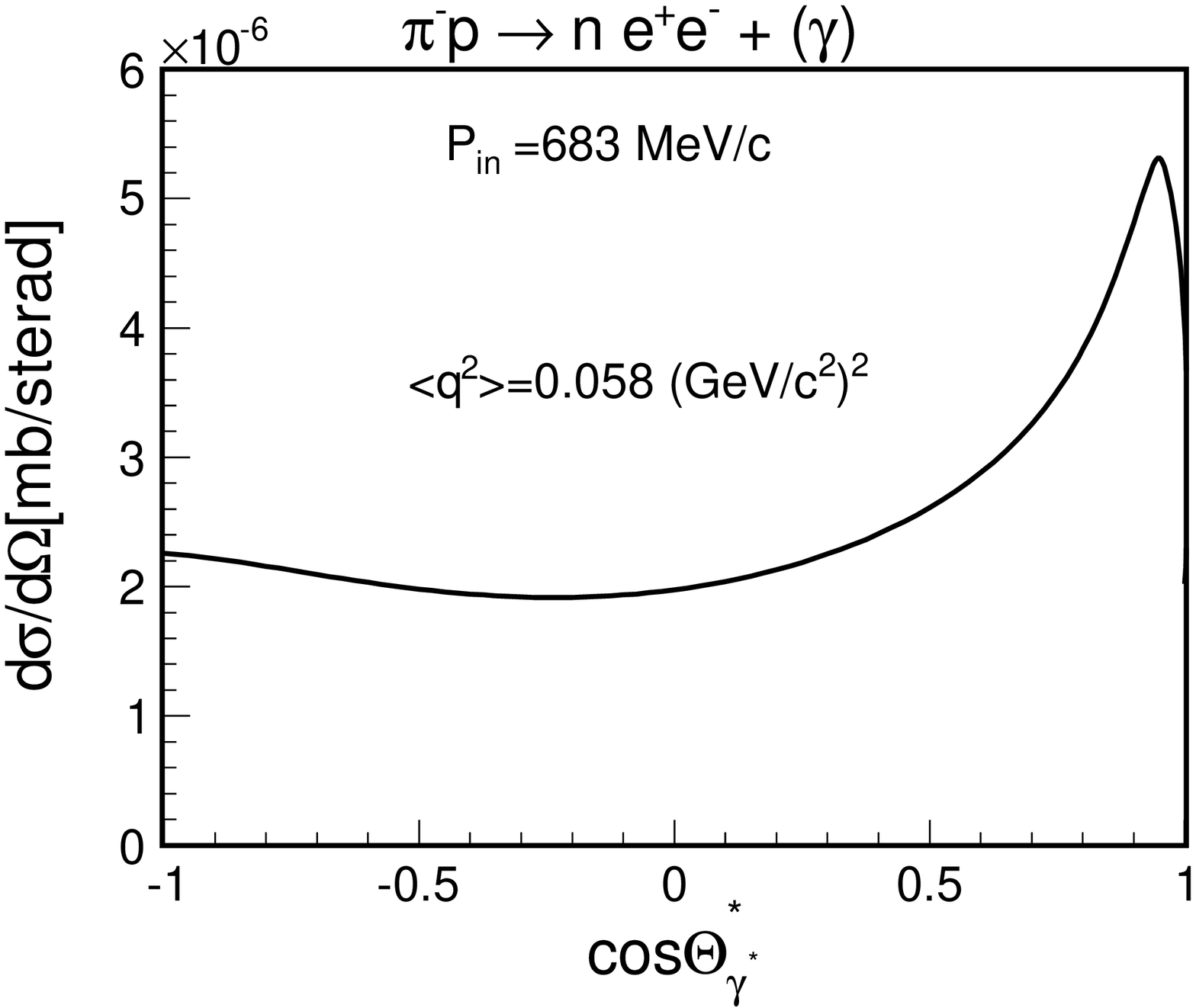,height=7.cm,width=7.2cm  }}
\caption{Left: angular distribution $d\sigma/d\cos\theta_{\gamma^*}$ for $\pi^- p\rightarrow e^+e^- n$
at initial pion momentum $P_{\pi}=$ 300 MeV/c \cite{JL:2018} and $<q^2>=$ 0.058 (GeV/c)$^2$.
where $\theta_{\gamma^*}$ is the angle of the virtual photon $\gamma^*$ . 
The solid curve corresponds to our calculations \cite{JL:2018}, the experimental
data are taken from \cite{Berezhnev:1976}.
Right: the same angle distribution, as the one presented to the left but at $P_{\pi}=$ 683 MeV/c.
}
\label{Fig.8}
\end{figure}

In Fig.~(\ref{Fig.8}, left) the angular distribution $d\sigma/d\cos\theta_{\gamma^*}$ is presented for reaction
$\pi^- p\rightarrow e^+e^- n$ at $P_{in}=$300 MeV/c 
and $0.046(GeV/c)^2\ge q^2\le 0.065(GeV/c)^2$ obtained in \cite{JL:2018}. The experimental data 
are taken from \cite{Berezhnev:1976}.
The solid curve corresponds to our calculations, e.g., the coherent sum of the one-nucleon and 
$\Delta$-exchange graphs with the positive phase sign of the second graph.
In Fig.~(\ref{Fig.8}, right) the similar angular distribution is presented for $P_{in}=$683 MeV/c
and the same interval of $q^2$. 
We present Fig.~(\ref{Fig.8}) to illustrate the similarity of such distribution\textcolor{red}{s}
at different initial momenta. 

\section{Conclusion}
In this paper we \textcolor{black}{have continued to analyze inverse} pion
electroproduction (IPE) processes $\pi p\rightarrow e^+e^- n$
\textcolor{black}{at intermediate} energies considered in \cite{JL:2018}. In
addition to that we have calculated the contributions of channels $\pi^- p \rightarrow \pi^0 n\rightarrow e^+e^-\gamma n$,
 $\pi^- p \rightarrow \eta^0 n\rightarrow e^+e^-\gamma n$ and  $\pi^- p \rightarrow \rho^0 n\rightarrow e^+e^- n$
to the effective mass distribution of dielectrons produced in the $\pi^- p$ process including the
final photon $\gamma$, which is not detected  
at the HADES facility. Therefore, we analyzed the process $\pi^- p \rightarrow \pi^0 n\rightarrow e^+e^- n (\gamma)$ and
investigated the sensitivity of observables, namely, the $e^+e^-$ effective mass distribution
$d\sigma/dM_{e^+e^-}$ and the angular distribution 
$d\sigma/d\cos\theta_{\gamma^*}$ to the electro-magnetic form factor $FF$ given by Eq.~(\ref{def:exponFF}). We found that
the inclusion of this form factor by calculation of the effective mass distribution of
the $e^+e^-$ pair can result in an enhancement in this spectrum at 
$M_{ee}>$ 300 MeV$/$c$^2$, which could be due to the electromagnetic properties of the time-like baryon 
transition. 
It can be verified by incoming HADES experiment.

{\bf  Acknowledgement\textcolor{black}{s}.}
We are grateful to T.Galatyuk,  
R.Holzman, V.P. Ladygin, D.Nitt, G.Pontecorvo, V.Pechenov, B.Ramstein, A.Rustamov, 
P.Salabura, J.Stroth for very useful discussions.

\section{Appendix: Parametrization of $\pi N \rightarrow \rho N$ ($\eta  N$) reactions}

  Within the Generalized Isobar Model (GIM)~\cite{GIM:1975,GIM:1992} 
$\pi N \rightarrow \pi \pi N$ 
reactions are described as quasi-two body reactions ($a + b \rightarrow c + d$):\\ 
\hspace*{1.5cm} $\pi N \rightarrow N^*(\Delta^*) \rightarrow N \rho^0$\\
\hspace*{1.5cm} $\pi N \rightarrow N^*(\Delta^*) \rightarrow N \eta^0$

with the subsequent decays:\\
\hspace*{2.5cm} $\rho \rightarrow e^+ e^-$,\\
\hspace*{2.5cm} $\eta^0 \rightarrow e^+e^-\gamma$.\\

 The parameters of the following resonances (****  and ***) were taken from 
 the Review~of~Particle~Properties:
$$
 \begin{array}{cc}
 N^*(1440) P11 & D^*(1600) P33\\
 N^*(1520) D13 & D^*(1620) S31\\
 N^*(1675) D15 & D^*(1700) D33\\
 N^*(1680) F15 & D^*(1900) S31\\
 N^*(1720) P13 & D^*(1905) F35\\
 N^*(2000) F15 & D^*(1910) P31\\
 N^*(2080) D13 & D^*(1920) P33\\
 N^*(2190) G17 & D^*(1940) D33\\
               & D^*(1950) F37
 \end{array}
$$                                         
     The spin and isospin relations were taken account.\\

  For quasi two-body reactions like $a + b \rightarrow c + d$ one can write
 $$d\sigma = \frac{1}{(2S_a+1)(2S_b+1)} \left(\frac{2\pi}{p}\right)^2 
 \sum_{\lambda_i}|<\lambda_d\lambda_c |T|\lambda_b\lambda_a>|^2 \times dPS\, ,$$
 $$<\lambda_d \lambda_c |T| \lambda_b \lambda_a> = \frac{1}{4\pi} \sum_j (2j+1)
   <\lambda_d \lambda_c |T_j| \lambda_b \lambda_a>  e^{i(\lambda - \mu) \varphi}
   d^j_{\lambda \mu}(\theta) \, .$$
where
 $\lambda=\lambda_a- \lambda_b$, $\mu=\lambda_c - \lambda_d$ are helicity 
 variables,\\
\hspace*{1.0cm} $d^j_{\lambda \mu}(\theta)$ is the rotation matrix,\\
\hspace*{1.0cm} $dPS$ is the phase space element.\\ 

  The angular distribution for the outgoing mesons ($\rho$ and $\eta$) in CMS 
is the following:
 $$\frac{d\sigma}{d\Omega} \sim const \times BW(\sqrt{s},M_R,\Gamma_R)$$

\end{document}